\documentclass[aps,showpacs,preprintnumbers,amsmath, amssymb]{revtex4}

\usepackage{float}
\usepackage{graphics,epsfig}
\usepackage{graphicx}
\usepackage{dcolumn}
\usepackage{bm}
\usepackage{booktabs}

\begin{document}
\baselineskip=0.8 cm
\title{{\bf Strong gravitational lensing in a squashed Kaluza-Klein G\"{o}del black hole}}
\author{Songbai Chen}
\email{csb3752@163.com}  \affiliation{Institute of Physics and
Department of Physics,
Hunan Normal University,  Changsha, Hunan 410081, P. R. China \\
Key Laboratory of Low Dimensional Quantum Structures and Quantum
Control (Hunan Normal University), Ministry of Education, P. R.
China.}

\author{Yue Liu}
\affiliation{Institute of Physics and Department of Physics,
Hunan Normal University,  Changsha, Hunan 410081, P. R. China \\
Key Laboratory of Low Dimensional Quantum Structures and Quantum
Control (Hunan Normal University), Ministry of Education, P. R.
China.}

\author{Jiliang Jing}
\email{jijing@hunnu.edu.cn} \affiliation{Institute of Physics and
Department of Physics,
Hunan Normal University,  Changsha, Hunan 410081, P. R. China \\
Key Laboratory of Low Dimensional Quantum Structures and Quantum
Control (Hunan Normal University), Ministry of Education, P. R.
China.}

\vspace*{0.2cm}
\begin{abstract}
\baselineskip=0.6 cm
\begin{center}
{\bf Abstract}
\end{center}

We investigate the strong gravitational lensing in a squashed
Kaluza-Klein black hole immersed in the G\"{o}del universe with
global rotation. Our result show that the strong gravitational
lensing in the squashed Kaluza-Klein G\"{o}del black hole spacetime
has some distinct behaviors from that in the Kerr case. In the
squashed Kaluza-Klein G\"{o}del black hole spacetime, the photon
sphere radius, the minimum impact parameter, the coefficient
$\bar{a}$, $\bar{b}$ and the deflection angle $\alpha(\theta)$ in
the $\phi$ direction are independent of whether the photon goes with
or against the global rotation of the G\"{o}del Universe. While in
the Kerr black hole, the values of these quantities for the prograde
photons are different from those for the retrograde photons.
Moreover, the coefficient of $\bar{b}$ increases with $j$ in the
squashed Kaluza-Klein G\"{o}del black hole, but decreases with $a$
in the Kerr case. We also probe the influence of the squashed effect
on the strong gravitational lensing in this black hole and find that
in the extremely squashed case $\rho_0=0$, the coefficient $\bar{a}$
is a constant $1$ and is independent of the global rotation of the
G\"{o}del Universe. Furthermore, we assume that the gravitational
field of the supermassive central object of the Galaxy can be
described by this metric and estimate the numerical values of the
coefficients and the main observables in the strong gravitational
lensing.

\end{abstract}

 \maketitle

\newpage
\section{Introduction}

It is well known that photons would be deviated from their straight
paths as they pass close to a compact and massive body.
Gravitational lensing is such a phenomena resulting from the
deflection of light rays in a gravitational field, which plays an
essential role in the astrophysics because it can help us extract
the information about the distant stars which are otherwise too dim
to be observed. The strong gravitational lensing is caused by a
compact object with a photon sphere (such as black
hole)\cite{Darwin}. When the photons pass close to the photon
sphere, the deflection angles of the light rays diverge and an
observer can detect two infinite sets of faint relativistic images
on each side of the black hole \cite{Vir,Vir1,Vir2}. It is shown
that the relativistic images can provide us not only some important
signatures about black holes in the Universe, but also profound
verification of alternative theories of gravity in their strong
field regime. Thus, the strong gravitational lensing by black holes
has been studied extensively in various theories of gravity
\cite{Vir3,Fritt,Bozza1,Eirc1,whisk,Bozza2,Bozza3,Bhad1,TSa1,AnAv,schen}.
Most studies of the strong gravitational lensing are focused on
black holes immersed in the rather idealized isotropic homogeneous
Universe.

It is more reasonable to consider the Universe background as
homogeneous but with global rotation since the rotation is a
universal phenomenon in our Universe. An original exact solution for
the rotating Universe was obtain by G\"{o}del through solving
Einstein equations with pressureless matter and negative
cosmological constant \cite{Godel}, which is a four-dimensional
spacetime and exhibits close timelike curves through every point.
The generalizations of G\"{o}del Universe in the minimal
five-dimensional gauged supergravity has been found in
\cite{Godel1,Godel12}. Like in the original four-dimensional
G\"{o}del solution, there also exist the close timelike curves in
these five-dimensional supersymmetric G\"{o}del spacetimes. Exact
solutions describing various black holes immersed in the
five-dimensional rotating G\"{o}del Universe were found in
\cite{Godelbh,Godelbh1,Godelbh2,Godelbh3,Godelbh4}, which obey to
the usual black hole thermodynamics. The studies indicate that the
quasinormal modes \cite{qnG} and Hawking radiation \cite{hakG} of
these black holes are considerably affected by the rotating
cosmological background. Using the squashing transformation, the new
squashed Kaluza-Klein  black hole solutions in the rotating universe
have been obtained in \cite{sqt1,sqt2}. In \cite{sqq1}, the wave
dynamics of the charged squashed Kaluza-Klein  G\"{o}del black hole
has been investigated, which shows that the squashed effect and the
cosmological rotational effect changes behavior of quasinormal
modes. The main purpose of this paper is to study the strong
gravitational lensing in the neutral squashed Kaluza-Klein
G\"{o}del black hole and to see what effects of the cosmological
rotation and compactness of the extra dimension on the coefficients
and the observables of gravitational lensing in the strong field
limit.

The plan of our paper is organized as follows. In Sec.II we
introduce briefly the neutral squashed Kaluza-Klein  black hole
immersed in the five-dimensional rotating G\"{o}del Universe. In
Sec.III we adopt to Bozza's method \cite{Bozza2,Bozza3,Bozza4} and
obtain the deflection angles for light rays propagating in the
squashed Kaluza-Klein  G\"{o}del black hole. In Sec.IV we suppose
that the gravitational field of the supermassive black hole at the
center of our Galaxy can be described by this metric and then obtain
the numerical results for the main observables in the strong
gravitational lensing. At last, we present a summary.

\section{The Squashed Kaluza-Klein G\"{odel} Black
Hole spacetime}

The static neutral squashed Kaluza-Klein G\"{odel} black hole is
described by the metric \cite{sqt1}
\begin{eqnarray}
ds^2&=&-f(r)dt^2+\frac{k(r)^2}{V(r)}dr^2-2g(r)\sigma_3dt+h(r)\sigma^2_3+
\frac{r^2}{4}[k(r)(\sigma^2_1+\sigma^2_2)+\sigma^2_3],\label{gm1}
\end{eqnarray}
with
\begin{eqnarray}
\sigma_1&=&\cos{\psi}d\theta+\sin{\psi}\sin{\theta}d\phi,\nonumber\\
\sigma_2&=&-\sin{\psi}d\theta+\cos{\psi}\sin{\theta}d\phi,\nonumber\\
\sigma_3&=&d\psi+\cos{\theta}d\phi.
\end{eqnarray}
Here coordinates $\theta\in [0,\pi)$,  $\phi\in [0,2\pi)$ and
$\psi\in [0,4\pi)$, and $r$ runs in the range $(0,\;r_{\infty})$.
The metric functions are
\begin{eqnarray}
f(r)&=&1-\frac{2M}{r^2},\;\;\;\;\;g(r)=jr^2,\;\;\;\;\;h(r)=-j^2r^2(r^2+2M),\nonumber\\
V(r)&=&1-\frac{2M}{r^2}+\frac{16j^2M^2}{r^2},\;\;\;\;\;\;k(r)=\frac{V(r_{\infty})r^4_{\infty}}{(r^2_{\infty}-r^2)^2}.
\end{eqnarray}
The parameter $M$ is the mass of the black hole, and $j$ denotes the
scale of the G\"{o}del background. The Killing horizon of the black
hole is given by the equation $V(r)=0$, so the radius of the black
hole horizon is $r^2_H=2M-16j^2M^2$. When $j=0$, Eq. (\ref{gm1})
reduces to the five-dimensional Kaluza-Klein  black hole with
squashed horizon \cite{IM}. As $r_{\infty} \rightarrow \infty $, one
has $k(r)\rightarrow1$, which means the squashing effect disappears
and the five-dimensional neutral black hole is recovered in the
G\"{o}del universe \cite{Godelbh}.

Using coordinate transformation
$\rho=\rho_0\frac{r^2}{r^2_{\infty}-r^2}$ and
$\tau=\sqrt{\frac{\rho_0}{\rho_0+\rho_M}}t$, we can find that the
metric (\ref{gm1}) can be rewritten as
\begin{eqnarray}
ds^2=-F(\rho)d\tau^2+\frac{K(\rho)}{G(\rho)}d\rho^2+\rho^2K(\rho)(d\theta^2+\sin^2\theta
d\phi^2)-2H(\rho)\sigma_3d\tau+D(\rho)\sigma^2_3,
\end{eqnarray}
where
\begin{eqnarray}
F(\rho)&=&1-\frac{\rho_M}{\rho},\;\;\;\;\;G(\rho)=1-\frac{\rho_H}{\rho},\;\;\;\;\;K(\rho)=1+\frac{\rho_0}{\rho},\nonumber\\
H(\rho)&=&j\frac{r^2_{\infty}}{K\rho_0}\sqrt{\rho_0(\rho_0+\rho_M)},\;\;\;\;
D(\rho)=\frac{r^2_{\infty}}{4K}- \frac{j^2\rho
r^4_{\infty}}{(\rho+\rho_0)^2(\rho_0+\rho_M)}\bigg[\rho(2\rho_M+\rho_0)+\rho_0\rho_M\bigg],
\end{eqnarray}
with
\begin{eqnarray}
\rho_M=\rho_0\frac{2M}{r^2_{\infty}-2M},\;\;\;\;\rho_H=\rho_0\frac{r^2_H}{r^2_{\infty}-r^2_H}.
\end{eqnarray}
Here $\rho_H$ is the radius of the black hole event horizon. The
parameter $\rho_0$ is a scale of transition from five-dimensional
spacetime to an effective four-dimensional one. The positive
parameters $r_{\infty}$, $\rho_0$ and $\rho_H$ are related by
$r^2_{\infty}=4\rho_0(\rho_0+\rho_H)$. The Komar mass of the black
hole is given by $M=\pi r_{\infty}\rho_M/G_5$ \cite{Ksq,Ksm}, where
$G_5$ is the five-dimensional gravitational constant. Since in the
squashed Kaluza-Klein black hole spacetime the relationship between
$G_5$ and $G_4$ ( the four-dimensional gravitational constant) can
be expressed as $G_5=2\pi r_{\infty}G_4$ \cite{Ksq,Ksm}, one can
obtain that the quantity $\rho_M=2G_4M$. The relation between
$\rho_H$ and $\rho_M$ is
\begin{eqnarray}
\rho_H=\frac{2(\rho_0+\rho_M)}{\sqrt{1+64j^2\rho^2_M}+1}-\rho_0.
\end{eqnarray}
It is clear that the radius of the black hole event horizon $\rho_H$
decreases with increase of the G\"{o}del parameter $j$. As $j$
vanishes, one can find that $\rho_H$ is coincide with $\rho_M$.

\section{Deflection angle in the Squashed Kaluza-Klein G\"{odel} Black
Hole spacetime}

In this section, we will study deflection angles of the light rays
when they pass close to a squashed Kaluza-Klein G\"{odel} black
hole, and then probe the effects of the G\"{odel} parameter $j$ and
the scale of extra dimension $\rho_0$ on the deflection angle and
the coefficients in the strong field limit. Here we consider only
the case the light ray is limited in the equatorial plane
$\theta=\frac{\pi}{2}$.  With this condition, we get the reduced
metric for the squashed Kaluza-Klein G\"{odel} black hole
\begin{eqnarray}
ds^2=-A(\rho)dt^2+B(\rho)d\rho^2+C(\rho)d\phi^2
+D(\rho)d\psi^2-2H(\rho)dtd\psi,\label{l1}
\end{eqnarray}
where
\begin{eqnarray}
A(\rho)=F(\rho),\;\;\;\;\;B(\rho)=\frac{K(\rho)}{G(\rho)},\;\;\;\;\;\;C(\rho)=\rho^2K(\rho).
\end{eqnarray}
For the null geodesics, we can obtain three constants of motion
\begin{eqnarray}
E&=-&g_{0\mu}\dot{x}^{\mu}=A(\rho)\dot{t}+H(\rho)\dot{\psi},\nonumber\\
L_{\phi}&=&g_{3\mu}\dot{x}^{\mu}=C(\rho)\dot{\phi},\nonumber\\
L_{\psi}&=&g_{4\mu}\dot{x}^{\mu}=D(\rho)\dot{\psi}-H(\rho)\dot{t}.
\end{eqnarray}
where a dot represents a derivative with respect to affine parameter
$\lambda$ along the geodesics. With these three constants, one can
find that the equations of motion of the photon can be simplified as
\begin{eqnarray}
\frac{dt}{d\lambda}&=&\frac{D(\rho)E-H(\rho)L_{\psi}}{H^2(\rho)+A(\rho)D(\rho)},\nonumber\\
\frac{d\phi}{d\lambda}&=&\frac{L_{\phi}}{C(\rho)},\nonumber\\
\frac{d\psi}{d\lambda}&=&\frac{H(\rho)E+A(\rho)L_{\psi}}{H^2(\rho)+A(\rho)D(\rho)}.
\end{eqnarray}
\begin{eqnarray}
\bigg(\frac{d\rho}{d\lambda}\bigg)^2
=\frac{1}{B(\rho)}\bigg[\frac{D(\rho)E-2H(\rho)EL_{\psi}-A(\rho)L^2_{\psi}}{H^2(\rho)+A(\rho)D(\rho)}-\frac{L^2_{\phi}}{C(\rho)}
\bigg].
\end{eqnarray}
From the $\theta$-component of the null geodesics in the equatorial
plane, we can obtain
\begin{eqnarray}
\frac{d\phi}{d\lambda}\bigg[D(\rho)\frac{d\psi}{d\lambda}-H(\rho)\frac{dt}{d\lambda}\bigg]=0,
\end{eqnarray}
which implies that either $\frac{d\phi}{d\lambda}=0$ or
$L_{\psi}=D(\rho)\frac{d\psi}{d\lambda}-H(\rho)\frac{dt}{d\lambda}=0$.
As done in the usual squashed Kaluza-Klein  black hole spacetime
\cite{schen}, here we set $L_{\psi}=0$, which implies that the total
angular momentum $J$ of the photo is equal to the constant
$L_{\phi}$. In doing so, one can get the effective potential for the
photon passing close to the black hole
\begin{eqnarray}
V(\rho)=\frac{1}{B(\rho)}\bigg[\frac{D(\rho)E}{H^2(\rho)+A(\rho)D(\rho)}-\frac{L^2_{\phi}}{C(\rho)}
\bigg].
\end{eqnarray}
Making use of this effective potential, one can obtain that the
impact parameter and the photon sphere equation are
\begin{eqnarray}
u=J=\sqrt{\frac{C(\rho)D(\rho)}{H(\rho)^2+A(\rho)D(\rho)}},\label{u1}
\end{eqnarray}
and
\begin{eqnarray}
D(\rho)\bigg[H(\rho)^2+A(\rho)D(\rho)\bigg]C'(\rho)-C(\rho)\bigg[D(\rho)^2A'(\rho)
+2D(\rho)H(\rho)H'(\rho)-H(\rho)^2D'(\rho)\bigg]=0,\label{phs-e}
\end{eqnarray}
respectively. Here we set $E=1$. The equations (\ref{u1}) and
(\ref{phs-e}) are more complex than those in the usual spherical
symmetric black hole spacetime.  As the G\"{o}del parameter
$j\rightarrow 0$, we find that the function $H(\rho)\rightarrow 0$,
which yields that the impact parameter (\ref{u1}) and the photon
sphere equation (\ref{phs-e}) reduce to those in the usual neutral
squashed Kaluza-Klein black hole spacetime \cite{schen}. The radius
of the photon sphere is the largest real root of Eq. (\ref{phs-e}),
which can be expressed as
\begin{eqnarray}
\rho_{ps}=\frac{-\mathcal{B}+\sqrt{\mathcal{B}^2-4\mathcal{A}\mathcal{C}}}{2\mathcal{A}},\label{phs1}
\end{eqnarray}
with
\begin{eqnarray}
\mathcal{A}&=&2\rho^2_M[1-
32j^2\rho_0(\rho_0+2\rho_M)+\sqrt{1+64j^2\rho^2_M}\;],\nonumber\\
\mathcal{B}&=&-(3\rho^3_0+9\rho^2_0\rho_M+8\rho_0\rho^2_M+6\rho^3_M)-32j^2\rho^2_0\rho^2_M(3\rho_0
+ 7\rho_M)+\rho_0(3\rho^2_0+ 9 \rho_0 \rho_M + 10 \rho^2_M)
\sqrt{1+64j^2\rho^2_M},\nonumber\\
\mathcal{C}&=&-2\rho_0\rho_M[2\rho^2_M+\rho^2_0+2\rho_0\rho_M+32j^2\rho^2_0
\rho^2_M-\rho_0(\rho_0+2\rho_M)\sqrt{1+64j^2\rho^2_M}\;].\label{phs111}
\end{eqnarray}
Obviously, it depends on both the G\"{o}del parameter $j$ and the
scale of transition $\rho_0$. As $j\rightarrow 0$, one can get
$\rho_{ps}=\frac{3\rho_M-\rho_0+\sqrt{\rho_0^2+10\rho_0\rho_M+9\rho^2_M}}{4}$,
which is consistent with that in the usual squashed Kaluza-Klein
black hole spacetime \cite{schen}. As $\rho_0$ approaches zero, the
radius of the photon sphere becomes
$\rho_{ps}=\frac{3\rho_M}{1+\sqrt{1+64j^2\rho^2_M}}$, which
decreases with the G\"{o}del parameter $j$. In Fig.(1), we set
$\rho_M=1$ and plot the variety of  the radius of the photon sphere
$\rho_{ps}$ with the parameters $j$ and $\rho_0$. It shows that with
the increase of $\rho_0$, $\rho_{ps}$ increases for the smaller $j$
and decreases for the larger $j$. For fixed $\rho_0$, it is easy to
obtain that $\rho_{ps}$ decreases monotonically with the increase of
the G\"{o}del parameter $j$. Moreover, it is well known that in the
Kerr black hole, the photon sphere radiuses are different for the
photons winding in the same direction (i.e., $a>0$) or in the
converse direction  (i.e., $a<0$) of the black hole rotation
\cite{Bozza3,Bozza4,GS}. However, one can obtain that in the
squashed Kaluza-Klein G\"{odel} black hole spacetime the photon
sphere radius $\rho_{ps}$ is independent of the sign of the
G\"{o}del parameter $j$ since all of quantities $\mathcal{A}$,
$\mathcal{B}$ and $\mathcal{C}$ in Eqs.(\ref{phs1}) and
(\ref{phs111}) are the function of $j^2$. Thus, the photon sphere
radius in the black hole spacetime keeps the same whether the photon
moves in the same or converse direction of the global rotation of
the G\"{o}del Universe. This means that the effects of the global
rotation parameter $j$ of the G\"{o}del Universe background on the
gravitational lensing is different from that of the rotation
parameter $a$ of the black hole itself.
\begin{figure}[ht]
\begin{center}
\includegraphics[width=6.1cm]{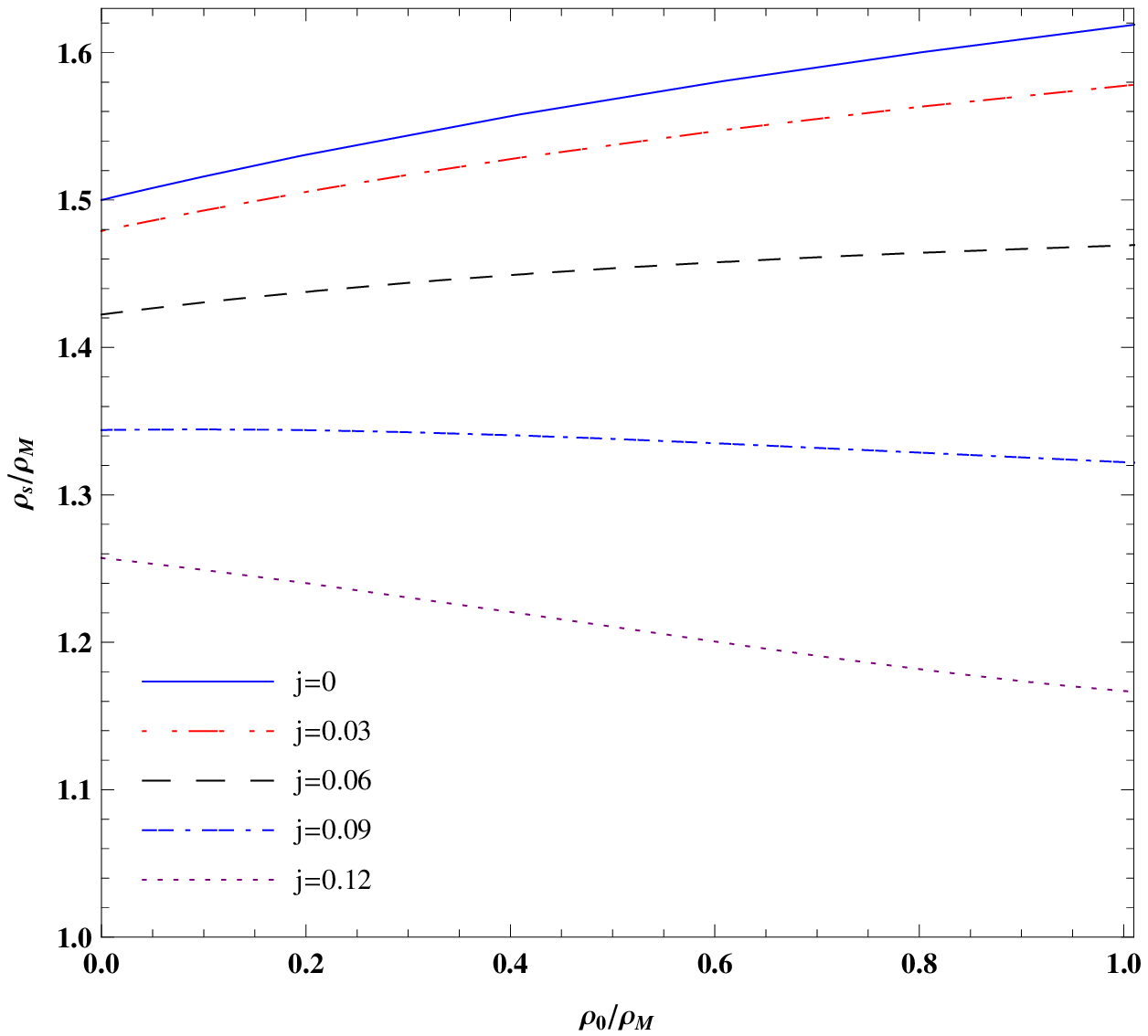}\;\;\;\includegraphics[width=6cm]{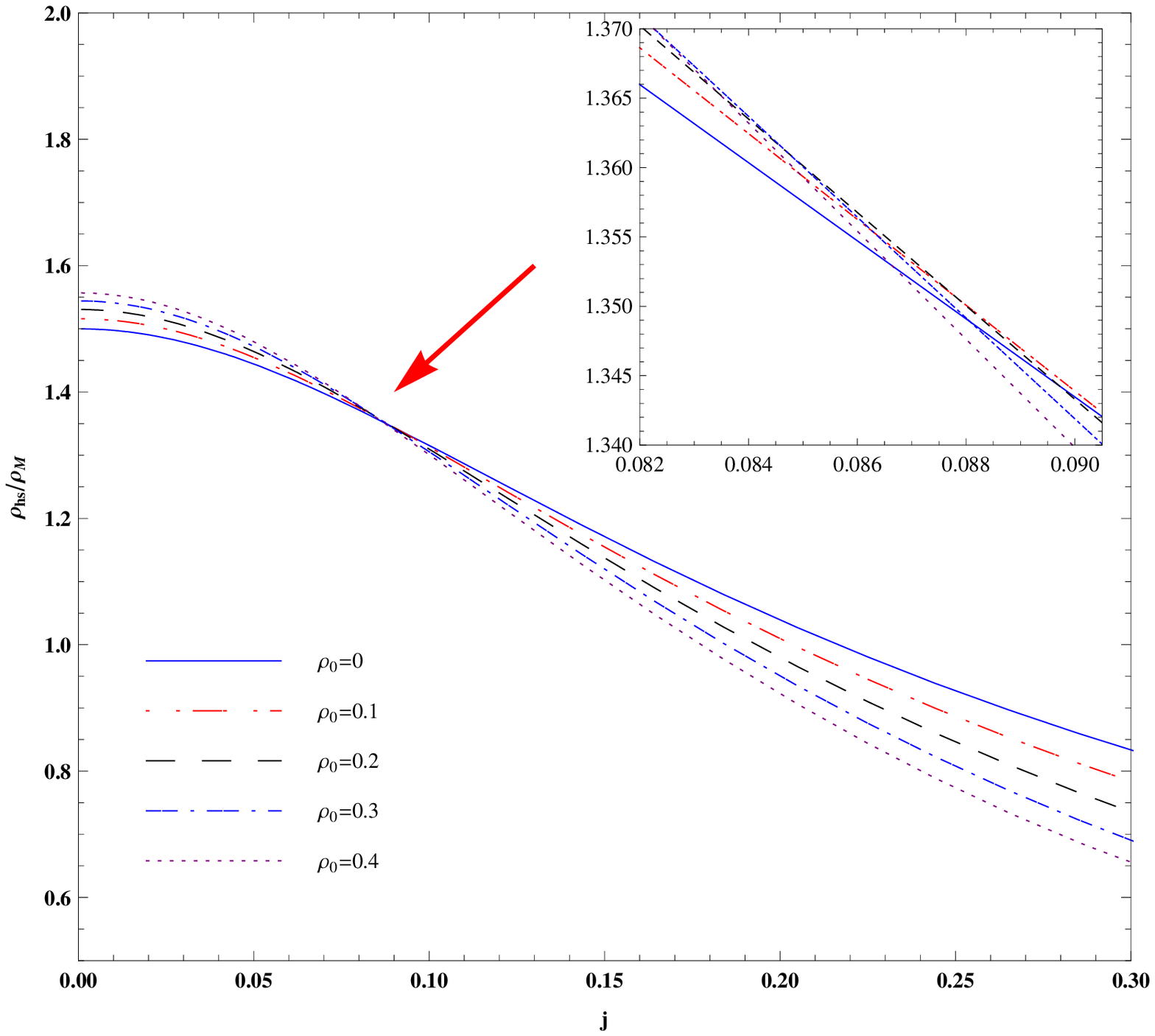}
\caption{Variety of the quantity $\rho_{ps}/\rho_M$ with
$\rho_0/\rho_M$ and $j$ in the squashed Kaluza-Klein G\"{o}del black
hole spacetime.}
\end{center}
\end{figure}

In the squashed Kaluza-Klein G\"{o}del black hole spacetime, the
deflection angles of $\phi$ and $\psi$ for the photon coming from
infinite are
\begin{eqnarray}
\alpha_{\phi}(\rho_{s})&=&I_{\phi}(\rho_s)-\pi,\label{dphi}\\
\alpha_{\psi}(\rho_{s})&=&I_{\psi}(\rho_s)-\pi,\label{dpsi}
\end{eqnarray}
respectively. Here $\rho_s$ is the closest approach distance,
$I_{\phi}(\rho_s)$ and $I_{\psi}(\rho_s)$ are
\begin{eqnarray}
I_{\phi}(\rho_s) &=&2\int^{\infty}_{\rho_s}\frac{\sqrt{B(\rho)
\mathcal{F}(\rho)C(\rho_s)}}{C(\rho)}\frac{1}{\sqrt{\mathcal{F}(\rho_s)-
\frac{\mathcal{F}(\rho)C(\rho_s)}{C(\rho)}}}d\rho,\label{int1}\\
I_{\psi}(\rho_s)&=&2\int^{\infty}_{\rho_s}\frac{H(\rho)}{D(\rho)}\sqrt{\frac{B(\rho)\mathcal{F}(\rho_s)}{\mathcal{F}(\rho)}}\frac{1}{\sqrt{\mathcal{F}(\rho_s)-
\frac{\mathcal{F}(\rho)C(\rho_s)}{C(\rho)}}}d\rho,\label{int2}
\end{eqnarray}
with
\begin{eqnarray}
\mathcal{F}(\rho)=\frac{H^2(\rho)+A(\rho)D(\rho)}{D(\rho)}.
\end{eqnarray}
As in the usual black hole spacetime, both of the deflection angles
increase when parameter $\rho_s$ decreases. If $\rho_s$ is equal to
the radius of the photon sphere $\rho_{ps}$, one can find that both
of the deflection angles diverge, which means that the photon is
captured by the black hole in this case. Moreover, from
Eq.(\ref{int1}) and (\ref{int2}), one can find that in the squashed
Kaluza-Klein G\"{o}del black hole spacetime both of the deflection
angles depend on the parameters $j$ and $\rho_0$, which implies that
we could in theory detect the global rotation of the universe and
the extra dimension through the gravitational lens. It is
interesting to note that the integral $I_{\phi}(\rho_s)$  is
function of $j^2$, which yields that the deflection angle
$\alpha_{\phi}(\rho_{s})$ is independent of whether the photon goes
with or against the global rotation of the G\"{o}del Universe.
However, from Eq. (\ref{int2}), we find that the integral
$I_{\psi}(\rho_s)$ contains the factor $j$, and then the deflection
angle $\alpha_{\psi}(\rho_{s})$ for the photon traveling in the same
direction as the global rotation of the G\"{o}del Universe is
different from that traveling in converse direction, which is
similar to that of the deflection angle in the $\phi$ coordinate in
the four-dimensional Kerr black hole spacetime \cite{Bozza3,Bozza4}.
The main reason is that in the equatorial plane
$\theta=\frac{\pi}{2}$ the global rotation of the spacetime is in
the $\psi$ direction rather than in the $\phi$ direction. As $j$
vanishes, one can find that the deflection angle of $\psi$ tends to
zero since $H(\rho)=0$, which reduces to that of in the usual
squashed Kaluza-Klein black hole spacetime \cite{schen}.

Here we will limit our attention to investigate the deflection angle
in the $\phi$ direction for the light rays passing close to the
black hole in the equatorial plane since it can be observed by
astronomical experiments. Moreover, it is very convenient for us to
compare with the results obtained in the usual black hole spacetime.

Defining a variable $z=1-\frac{\rho_s}{\rho}$, one can rewrite
Eq.(\ref{int1}) as \cite{Bozza2,Bozza3}
\begin{eqnarray}
I_{\phi}(\rho_s)&=&\int^{1}_{0}R(z,\rho_s)f(z,\rho_s)dz,\label{in1}
\end{eqnarray}
with
\begin{eqnarray}
R(z,\rho_s)&=&2\frac{\rho^2}{\rho_sC(\rho)}\sqrt{B(\rho)
\mathcal{F}(\rho)C(\rho_s)}\nonumber\\
&=&2\bigg\{\frac{\rho^2_M(\rho_s+\rho_0)(1+\sqrt{1+64j^2\rho^2_M})}
{2\rho^2_M[\rho_s+\rho_0(1-z)]-\rho_0(\sqrt{1+64j^2\rho^2_M}-1)[(1-z)\rho_0\rho_M+\rho_s(2\rho_M+\rho_0)]}\bigg\}^{\frac{1}{2}},
\end{eqnarray}
and
\begin{eqnarray}
f(z,\rho_s)&=&\frac{1}{\sqrt{\mathcal{F}(\rho_s)-\mathcal{F}(\rho)C(\rho_s)/C(\rho)}}.\label{ffq}
\end{eqnarray}
\begin{figure}[ht]
\begin{center}
\includegraphics[width=6cm]{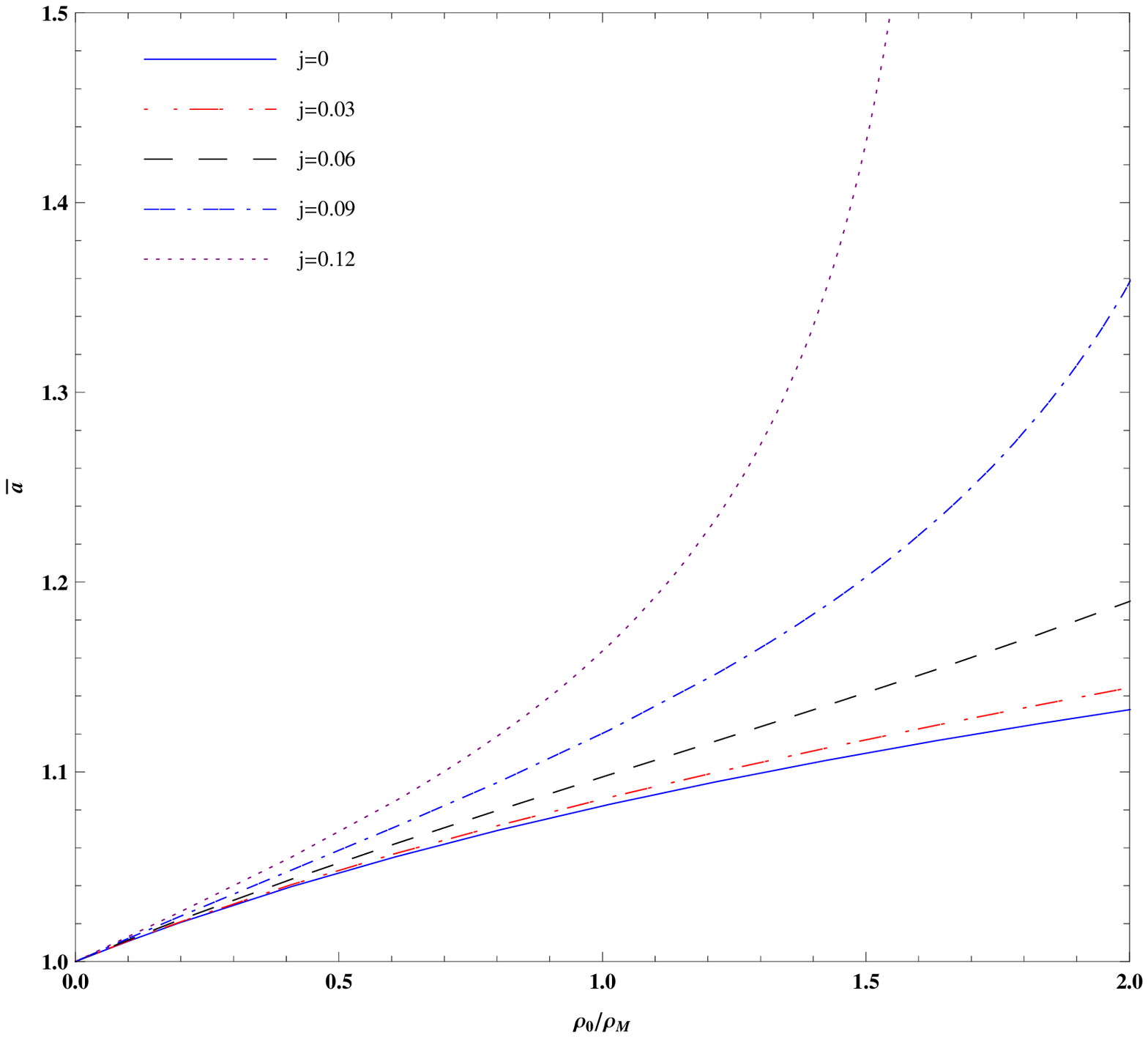}\;\;\;\includegraphics[width=6cm]{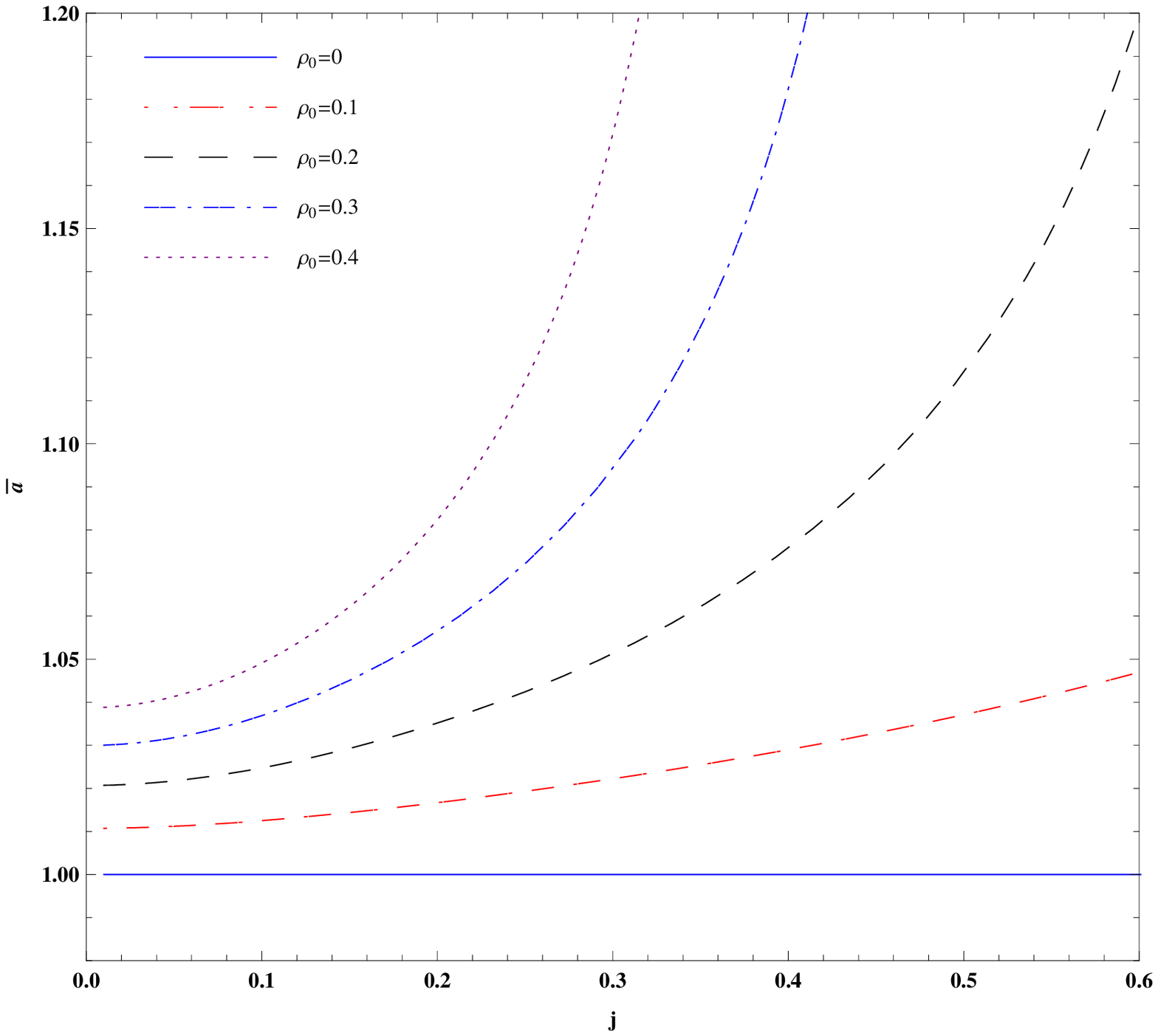}
\caption{Variety of the coefficient $\bar{a}$ with $\rho_0/\rho_M$
and $j$ in the squashed Kaluza-Klein G\"{o}del black hole
spacetime.}
\end{center}
\end{figure}
Here we consider only the small $j$ case since the small rotation of
the G\"{o}del cosmological background seems the most reasonable in
phenomenology. Moreover, the function $R(z, \rho_s)$ in the small
$j$ case is regular for all values of $z$ and $\rho_s$. From
Eq.(\ref{ffq}), one can find that the function $f(z, \rho_s)$
diverges as $z$ tends to zero. Thus, the integral (\ref{in1}) can be
split into the divergent part $I_D(\rho_s)$ and the regular one
$I_R(\rho_s)$
\begin{eqnarray}
I_D(\rho_s)&=&\int^{1}_{0}R(0,\rho_{ps})f_0(z,\rho_s)dz, \nonumber\\
I_R(\rho_s)&=&\int^{1}_{0}[R(z,\rho_s)f(z,\rho_s)-R(0,\rho_{ps})f_0(z,\rho_s)]dz
\label{intbr}.
\end{eqnarray}
Expanding the argument of the square root in $f(z,\rho_{s})$ to the
second order in $z$, we have
\begin{eqnarray}
f_s(z,\rho_{s})=\frac{1}{\sqrt{p(\rho_{s})z+q(\rho_{s})z^2}},
\end{eqnarray}
where
\begin{eqnarray}
p(\rho_{s})&=&\frac{\rho_s}{C(\rho_s)}\bigg[C'(\rho_s)\mathcal{F}(\rho_s)-C(\rho_s)\mathcal{F}'(\rho_s)\bigg],  \nonumber\\
q(\rho_{s})&=&\frac{\rho^2_s}{2C(\rho_s)}\bigg[2C'(\rho_s)C(\rho_s)\mathcal{F}'(\rho_s)-2C'(\rho_s)^2\mathcal{F}(\rho_s)
+\mathcal{F}(\rho_s)C(\rho_s)C''(\rho_s)-C^2(\rho_s)\mathcal{F}''(\rho_s)\bigg].
\end{eqnarray}
Obviously, as $\rho_{s}$ tends to $\rho_{ps}$, one can obtain easily
that the leading term of the divergence in $f_s(z,\rho_{s})$ is
$z^{-1}$ since the coefficient $p(\rho_{s})$ approaches zero, which
implies that the integral (\ref{in1}) diverges logarithmically.
Thus, the deflection angle in the $\phi$ direction in the strong
field region can be approximated very well as \cite{Bozza2}
\begin{eqnarray}
\alpha(\theta)=-\bar{a}\log{\bigg(\frac{\theta
D_{OL}}{u_{ps}}-1\bigg)}+\bar{b}+O(u-u_{ps}), \label{alf1}
\end{eqnarray}
\begin{figure}[ht]
\begin{center}
\includegraphics[width=6cm]{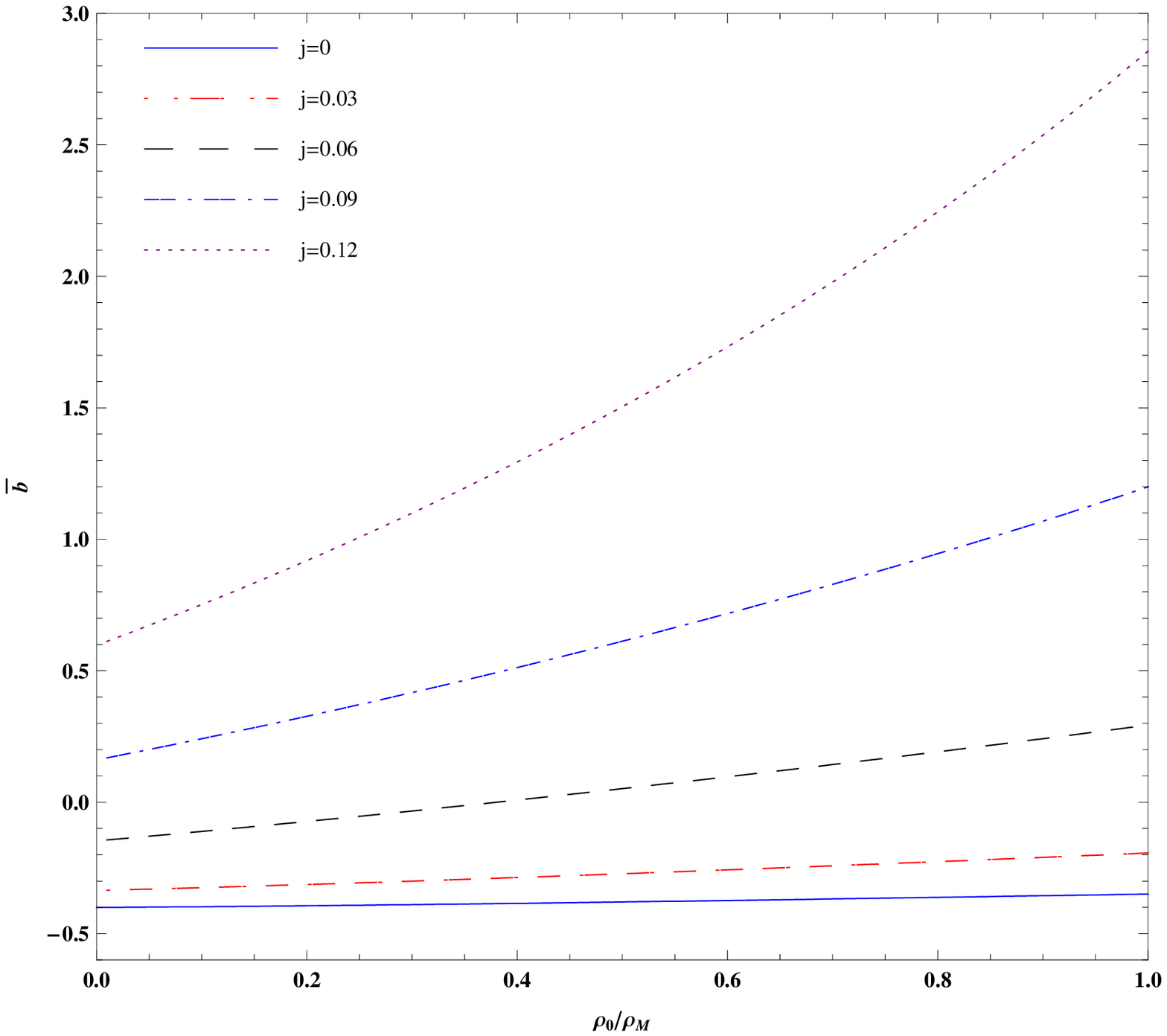}\;\;\;\includegraphics[width=6cm]{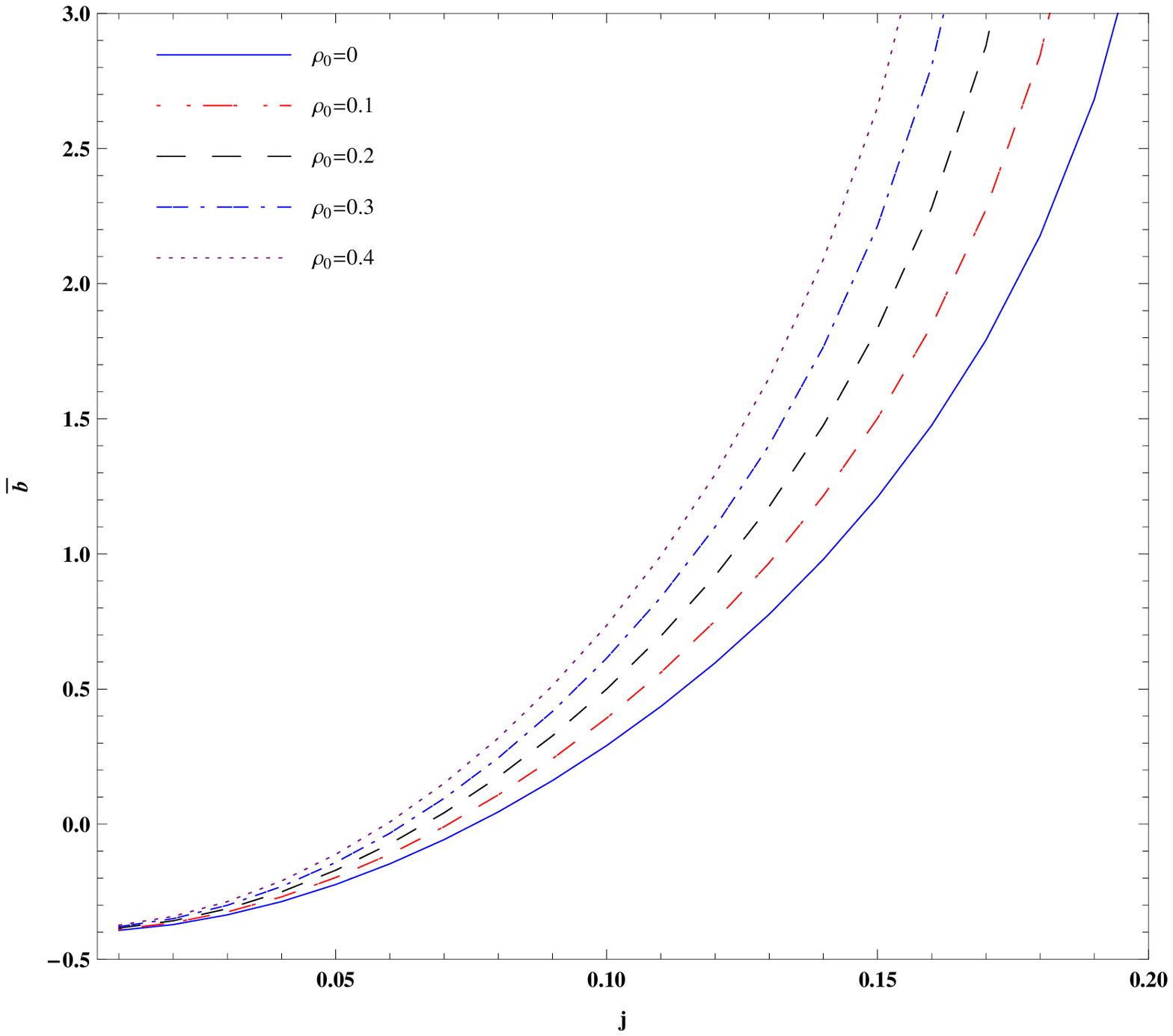}
\caption{Variety of the coefficient $\bar{b}$ with $\rho_0/\rho_M$
and $j$ in the squashed Kaluza-Klein G\"{o}del black hole
spacetime.}
\end{center}
\end{figure}
with
\begin{eqnarray}
&\bar{a}&=\frac{R(0,\rho_{ps})}{2\sqrt{q(\rho_{ps})}}, \nonumber\\
&\bar{b}&=
-\pi+b_R+\bar{a}\log{\frac{\rho^2_{hs}[C''(\rho_{ps})\mathcal{F}(\rho_{ps})-C(\rho_{ps})\mathcal{F}''(\rho_{ps})]}{u_{ps}
\sqrt{\mathcal{F}^3(\rho_{ps})C(\rho_{ps})}}}, \nonumber\\
&b_R&=I_R(\rho_{ps}), \;\;\;\;\;\;\;\;
u_{ps}=\sqrt{\frac{C(\rho_{ps})}{\mathcal{F}(\rho_{ps})}}.\label{coa1}
\end{eqnarray}
The quantity $D_{OL}$ is the distance between observer and
gravitational lens. Repeating the operations above, one can obtain a
similar strong gravitational lensing formula for the deflection
angle in the $\psi$ direction ( $\alpha_{\psi}(\theta)$) in which
the forms of coefficients are different from that of $\bar{a}$ and
$\bar{b}$ in Eq.(\ref{coa1}). As $\rho_{s}$ tends to $\rho_{ps}$, we
find that the deflection angle $\alpha_{\psi}(\theta)$ also diverges
logarithmically. However, $\alpha_{\psi}(\theta)$ cannot actually be
observed by astronomical experiments. So we do not consider it in
the following discussion.

Combining with Eqs.(\ref{phs1}), (\ref{alf1}) and (\ref{coa1}), we
can probe the properties of strong gravitational lensing in the
squashed Kaluza-Klein G\"{o}del black hole spacetime and explore the
effects of the rotation parameter of cosmological background $j$ on
the deflection angle in the strong field limit.  In the Figs.
(2)-(3), we plot the variations of the coefficients $\bar{a}$ and
$\bar{b}$ in the deflection angle (\ref{alf1}) with the parameters
$j$ and $\rho_0$. As $j$ tends to zero, these quantities reduce to
those in the squashed Kaluza-Klein black hole in the cosmological
background without rotation \cite{schen}. Moreover, we can see that
for fixed $j$, both of the coefficients $\bar{a}$ and $\bar{b}$
increase with the size of the extra dimension $\rho_0$, which are
similar to those in the usual squashed Kaluza-Klein black hole
spacetime. But these two coefficients increase more quickly than in
the case $j=0$. For fixed $\rho_0$, the coefficient $\bar{a}$ in
general increases with the increase of $j$. However, we also find
that in the extremely squashed case $\rho_0=0$,  $\bar{a}$ is a
constant $1$ and is independent of the G\"{o}del parameter $j$.
Fig.(3) tells us that the coefficient $\bar{b}$ increases with $j$
for arbitrary $\rho_0$, which is different from the change of the
coefficient $\bar{b}$ with the rotation parameter $a$ in Kerr black
hole spacetime.
\begin{figure}[ht]
\begin{center}
\includegraphics[width=6cm]{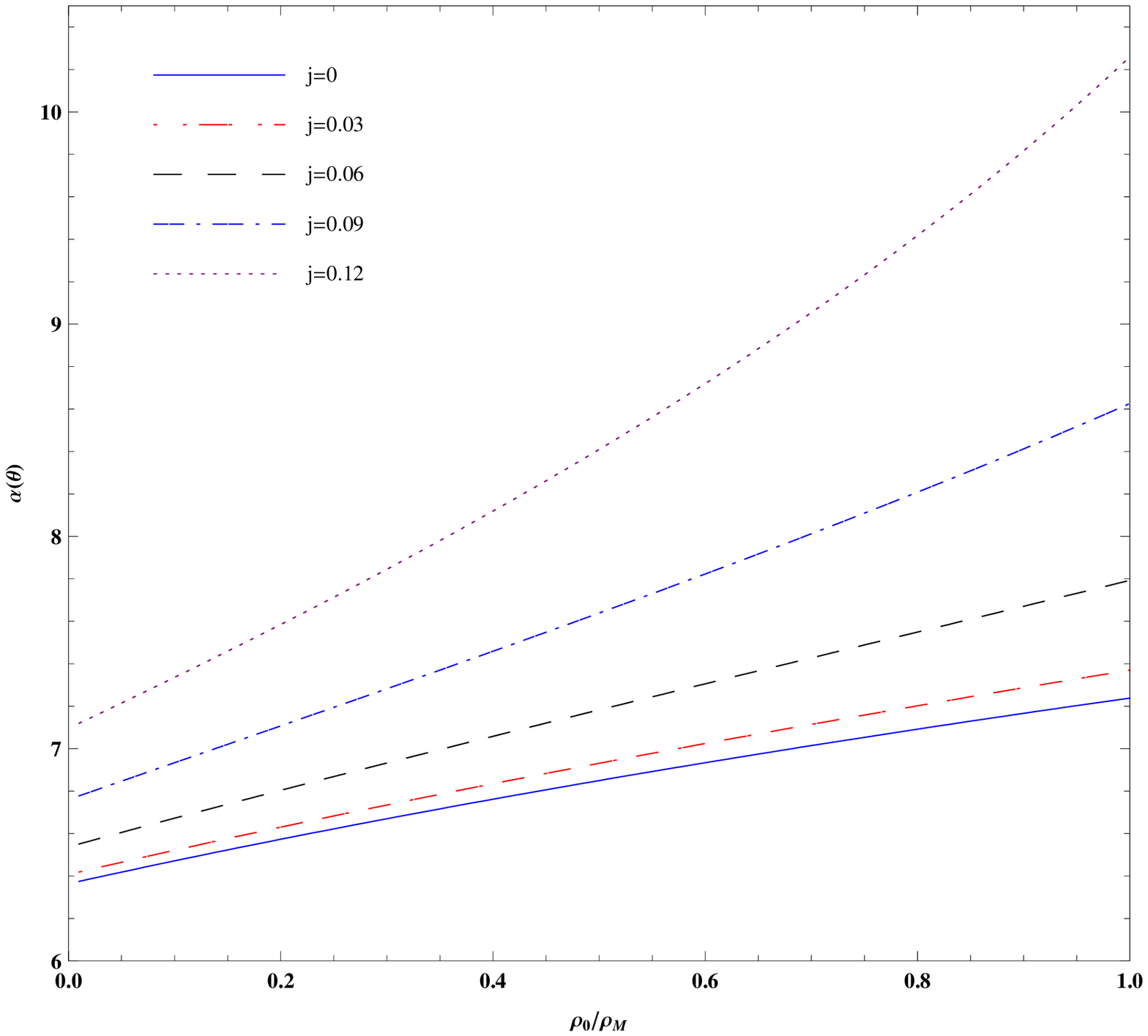}\;\;\;\includegraphics[width=6cm]{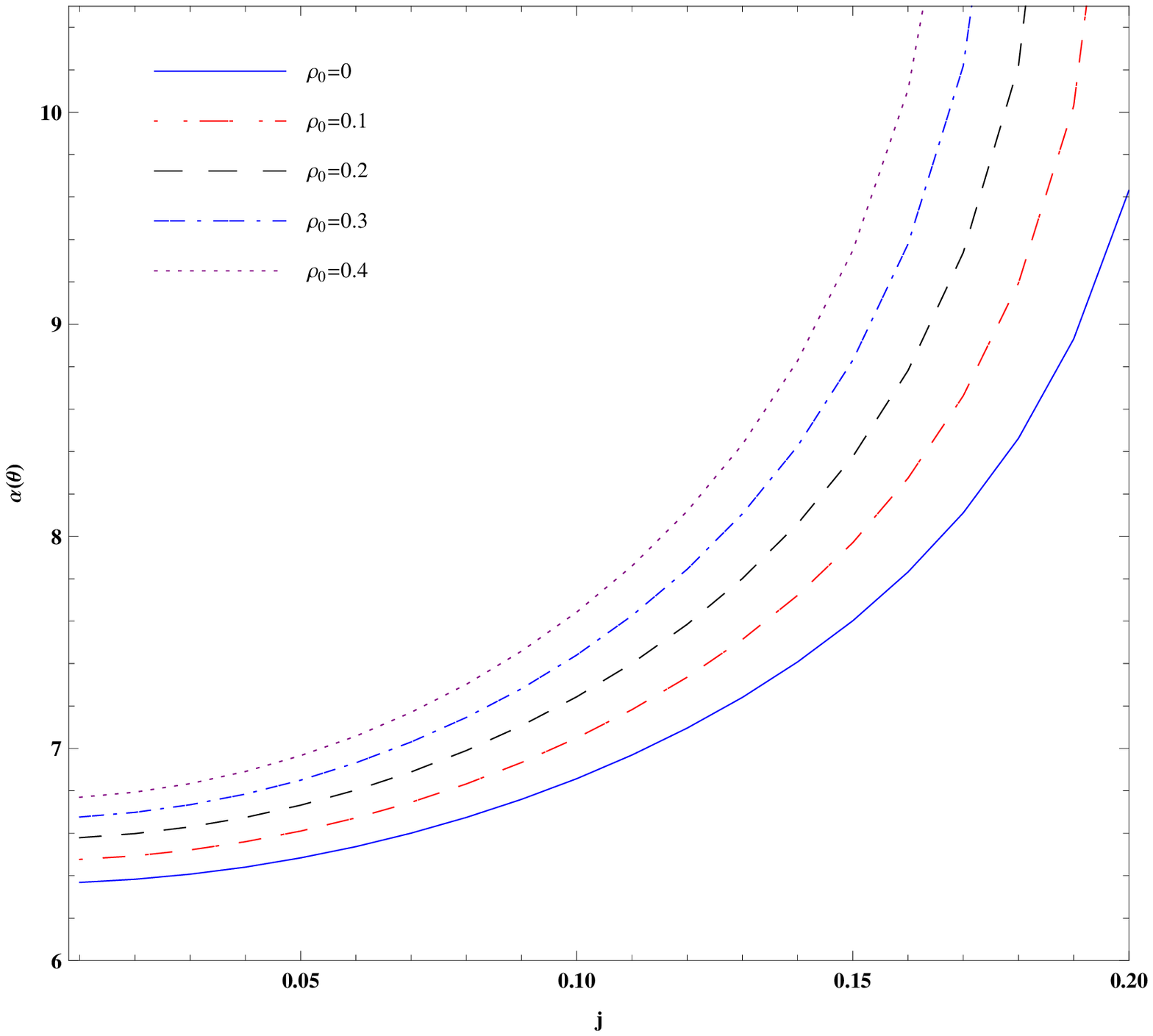}
\caption{Deflection angles in the squashed Kaluza-Klein G\"{o}del
black hole spacetime evaluated at $u=u_{ps}+0.003$ as functions of
$\rho_0/\rho_M$ and $j$.}
\end{center}
\end{figure}
In Fig.(4), We present the deflection angle $\alpha(\theta)$
evaluated at $u=u_{ps}+0.003$, which shows that the larger
parameters $j$ and $\rho_0$ lead to the bigger deflection angle
$\alpha(\theta)$ for the light propagated in the squashed
Kaluza-Klein G\"{o}del black hole spacetime \cite{schen}. Comparing
with those in the Schwarzschild black hole spacetime, we could
extract information about both the rotation of the cosmological
background and the size of the extra dimension by using strong field
gravitational lensing.

\section{Observational gravitational lensing parameters}

Let us now to see how the G\"{o}del parameter $j$ and the scale
parameter $\rho_0$ affect the observables in the strong
gravitational lensing. Assuming that the spacetime of the
supermassive black hole at the Galactic center of Milky Way can be
described by the squashed Kaluza-Klein G\"{o}del black hole metric,
we can estimate the numerical values for the coefficients and
observables of gravitational lensing in the strong field limit.

We consider the simplest geometric disposition when the source, the
lens and the observer are highly aligned so that the lens equation
in strong gravitational lensing can be simplified as \cite{Bozza1}
\begin{eqnarray}
\beta=\theta-\frac{D_{LS}}{D_{OS}}\Delta\alpha_{n},
\end{eqnarray}
where the quantity $D_{LS}$ denote the distance between the lens and
the source. $D_{OS}$ is the distance between the observer and the
source, which is related to $D_{LS}$ and  $D_{OS}$ by
$D_{OS}=D_{LS}+D_{OL}$ for this simplest geometry. $\beta$ and
$\theta$ are the angular separations between the source and the
lens, and between the imagine and the lens, respectively.
$\Delta\alpha_{n}=\alpha-2n\pi$ is the offset of deflection angle,
and $n$ is an integer. Since $u_{ps}\ll D_{OL}$, one can find that
the $n$-th image position $\theta_n$ and the $n$-th image
magnification $\mu_n$ can be approximated as
\begin{eqnarray}
\theta_n=\theta^0_n+\frac{u_{ps}(\beta-\theta^0_n)e^{\frac{\bar{b}-2n\pi}{\bar{a}}}D_{OS}}{\bar{a}D_{LS}D_{OL}},
\end{eqnarray}
\begin{eqnarray}
\mu_n=\frac{u^2_{hs}(1+e^{\frac{\bar{b}-2n\pi}{\bar{a}}})e^{\frac{\bar{b}-2n\pi}{\bar{a}}}D_{OS}}{\bar{a}\beta
D_{LS}D^2_{OL}},
\end{eqnarray}
respectively. Here $\theta^0_n$ is the image positions corresponding
to $\alpha=2n\pi$. In the limit $n\rightarrow \infty$, the minimum
impact parameter $u_{ps}$ is related to the asymptotic position of a
set of images $\theta_{\infty}$ by a simple form
\begin{eqnarray}
u_{ps}=D_{OL}\theta_{\infty}.\label{uhs1}
\end{eqnarray}
In order to obtain the coefficients $\bar{a}$ and $\bar{b}$, one
needs to separate the outermost image from all the others. As in
Refs.\cite{Bozza2,Bozza3},  we consider here the simplest situation
in which only the outermost image $\theta_1$ is resolved as a single
image and all the remaining ones are packed together at
$\theta_{\infty}$.  With these simplifications, one can find that
the angular separation between the first image and other ones $s$
and the ratio of the flux from the first image and those from the
all other images $\mathcal{R}$ can be expressed as
\cite{Bozza2,Bozza3}
\begin{eqnarray}
s&=&\theta_1-\theta_{\infty}=\theta_{\infty}e^{\frac{\bar{b}-2\pi}{\bar{a}}},\nonumber\\
\mathcal{R}&=&\frac{\mu_1}{\sum^{\infty}_{n=2}\mu_n}=e^{\frac{2\pi}{\bar{a}}}.\label{ss1}
\end{eqnarray}
Through measuring these simple observations $s$, $\mathcal{R}$, and
$\theta_{\infty}$, one can obtain the strong deflection limit
coefficients $\bar{a}$, $\bar{b}$ and the minimum impact parameter
$u_{ps}$. Comparing their values with those predicted by the
theoretical models, we can obtain the characteristics information
about of the lens object stored in the strong gravitational lensing.
\begin{figure}[ht]
\begin{center}
\includegraphics[width=6cm]{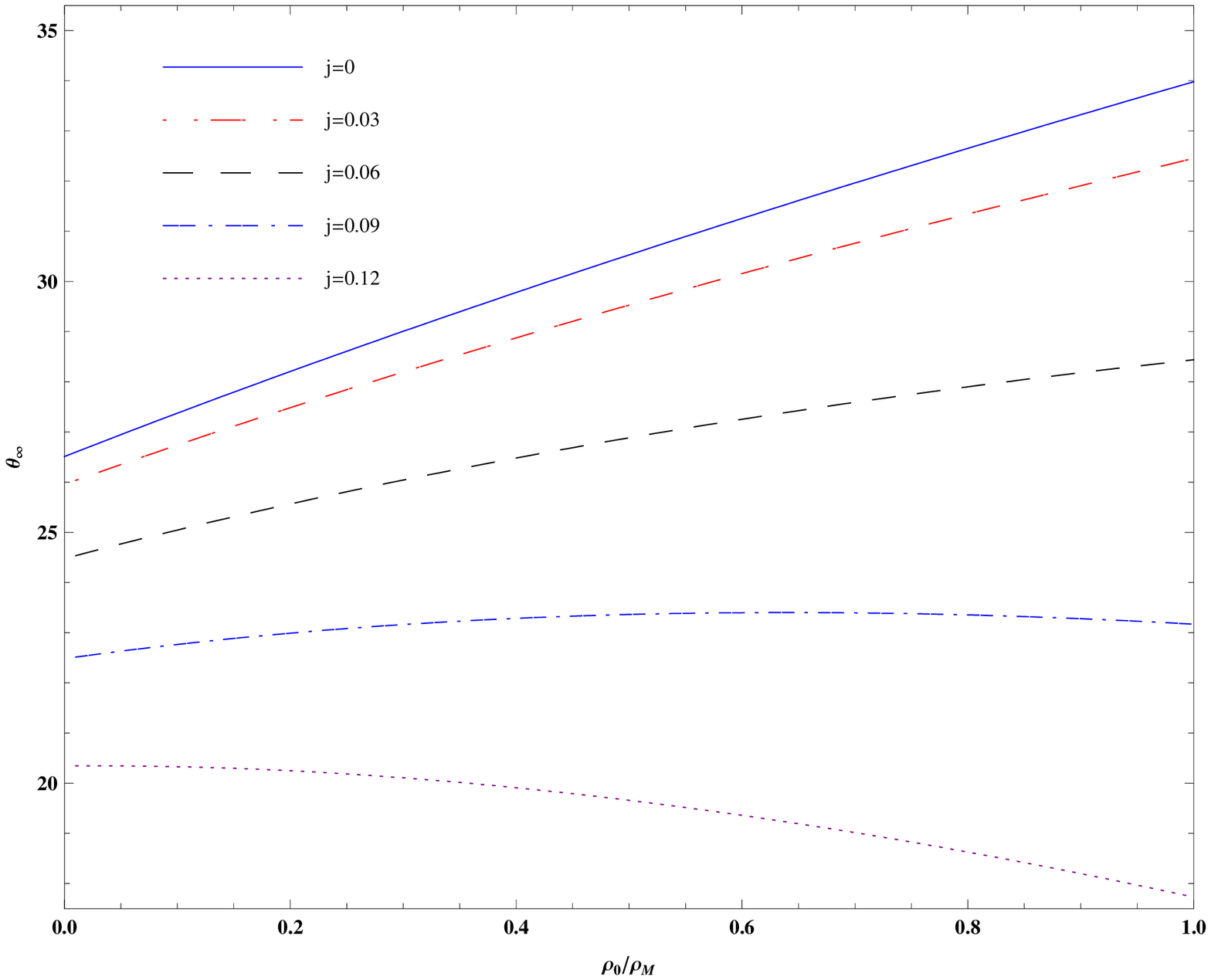}\;\;\;\includegraphics[width=6cm]{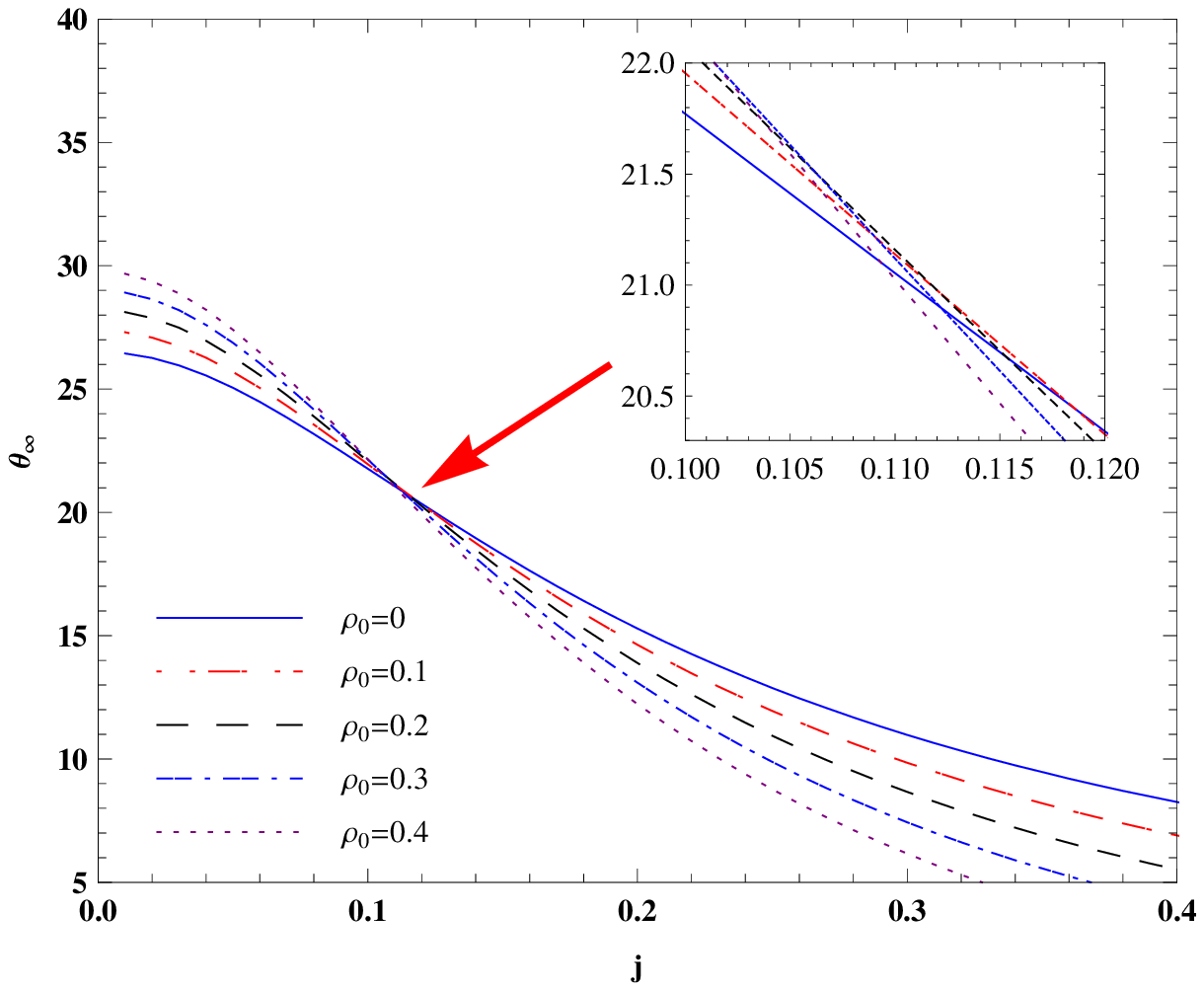}
\caption{Gravitational lensing by the Galactic center black hole.
Variation of the values of the angular position $\theta_{\infty}$
with parameters $\rho_0/\rho_M$ and $j$ in the squashed Kaluza-Klein
black hole spacetime.}
\end{center}
\end{figure}
\begin{figure}[ht]
\begin{center}
\includegraphics[width=6cm]{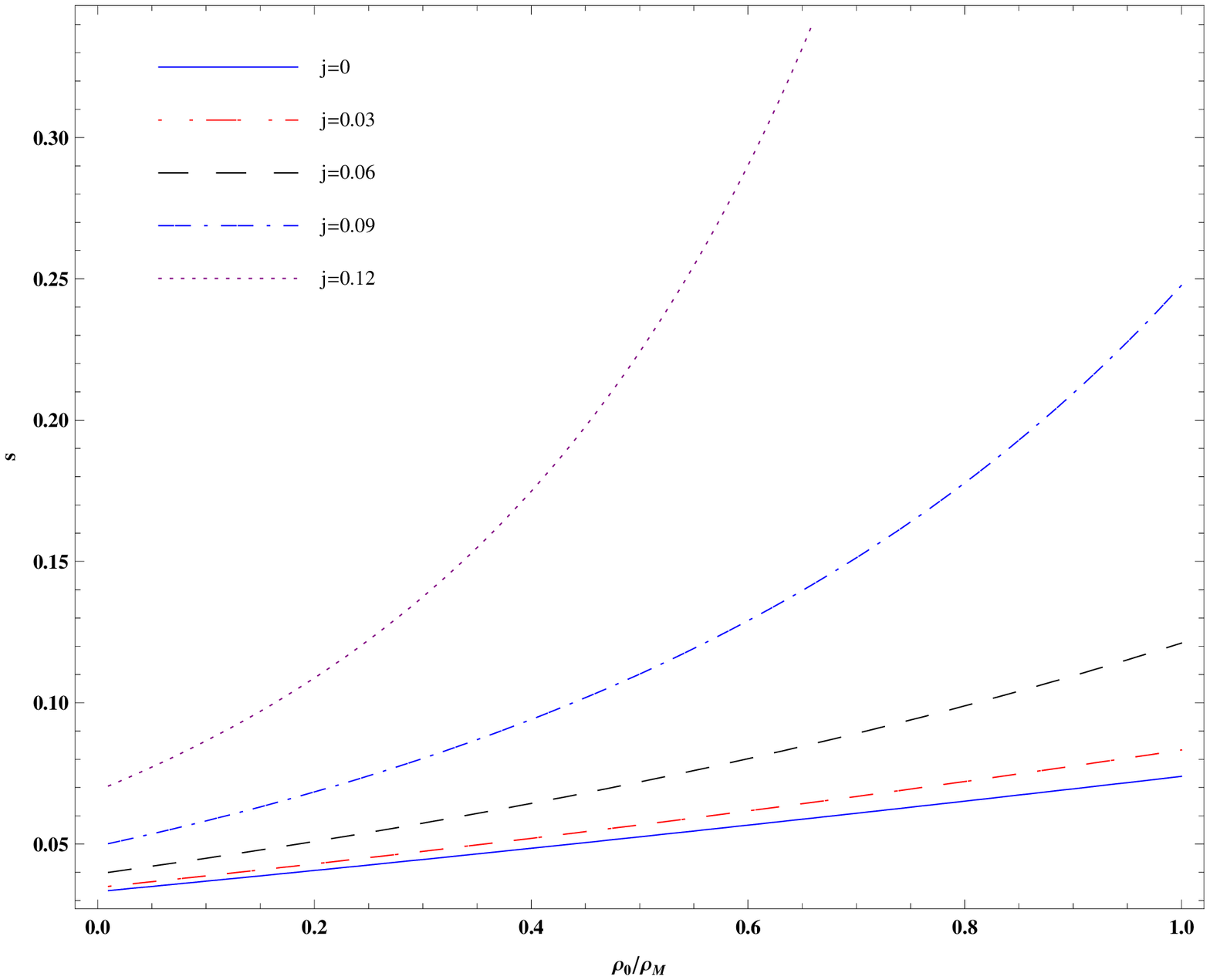}\;\;\;\includegraphics[width=6cm]{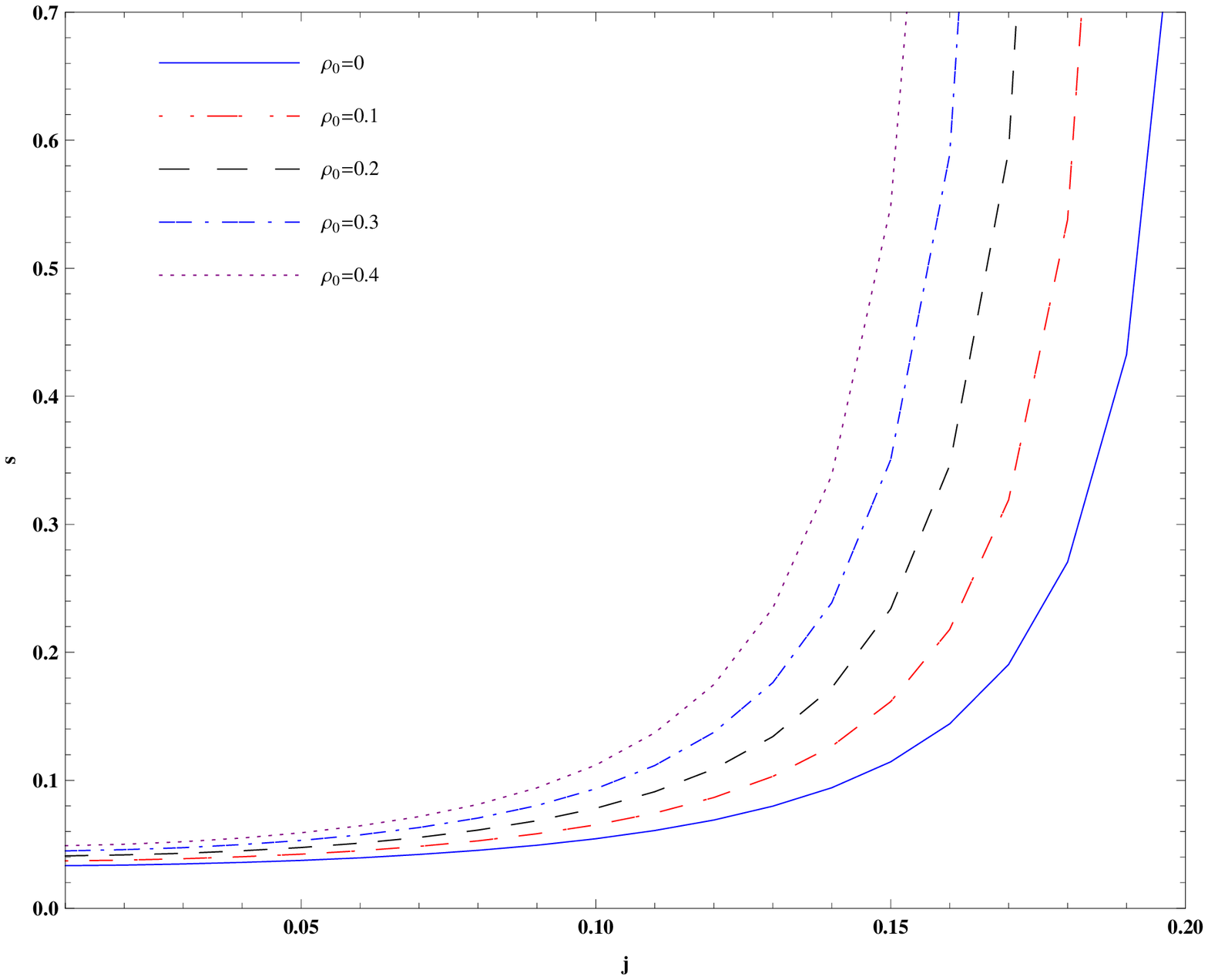}
\caption{Gravitational lensing by the Galactic center black hole.
Variation of the values of the angular separation $s$ with
parameters $\rho_0/\rho_M$ and $j$ in the squashed Kaluza-Klein
black hole spacetime.}
\end{center}
\end{figure}
\begin{figure}[ht]
\begin{center}
\includegraphics[width=6cm]{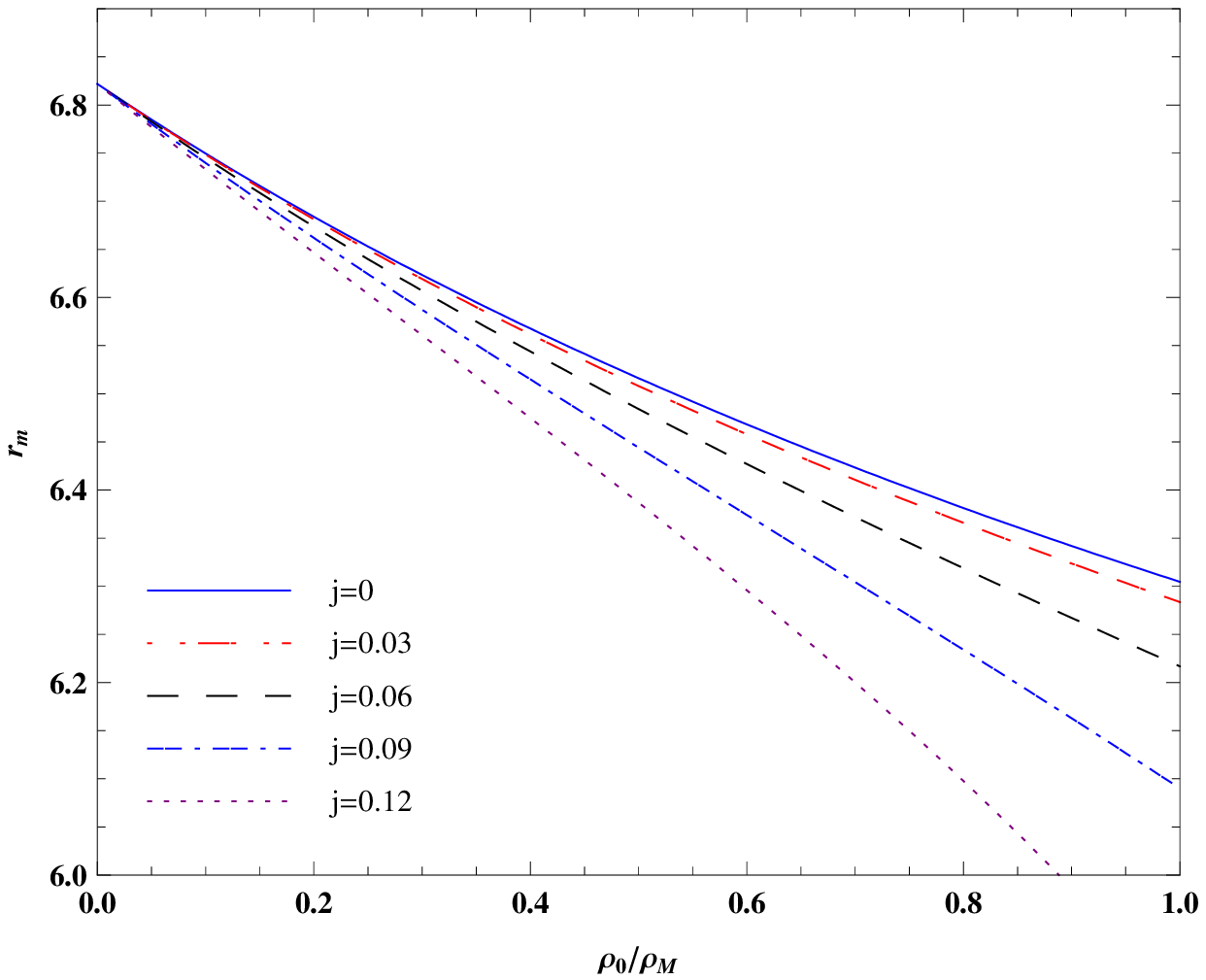}\;\;\;\includegraphics[width=6cm]{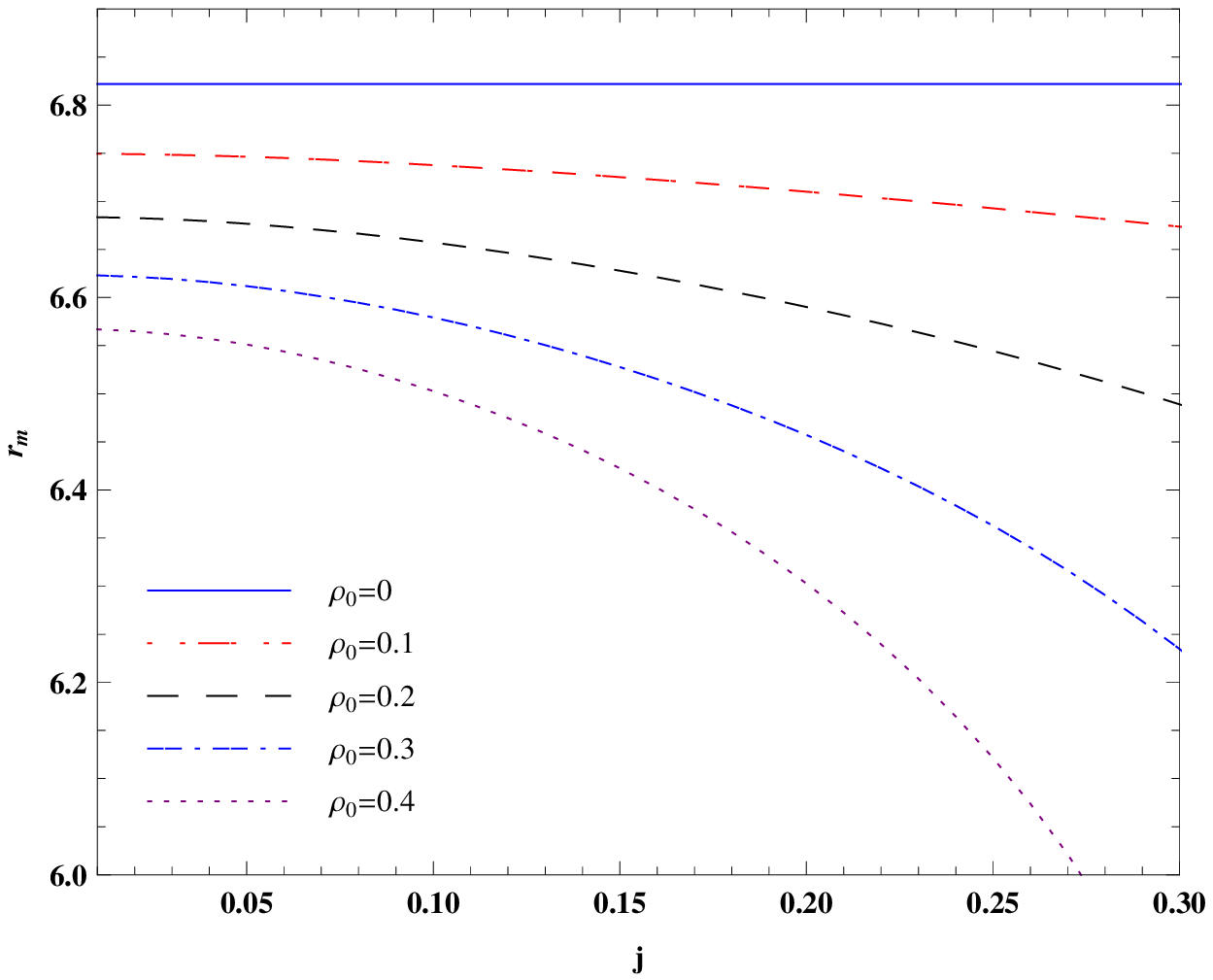}
\caption{Gravitational lensing by the Galactic center black hole.
Variation of the values of the relative magnitudes $r_m$ with
parameters $\rho_0/\rho_M$ and $j$ in the squashed Kaluza-Klein
black hole spacetime.}
\end{center}
\end{figure}

The mass of the central object of our Galaxy is estimated to be
$4.4\times 10^6M_{\odot}$  and its distance is around $8.5kpc$
recently, so the ratio of the mass to the distance $G_4M/D_{OL}
\approx2.4734\times10^{-11}$ \cite{grf}. Here $D_{OL}$ is the
distance between the lens and the observer in the $\rho$
coordination rather than that in $r$ coordination because that in
the five-dimensional spacetime the dimension of the black hole mass
$M$ is the square of that in the polar coordination $r$. With help
of  Eqs. (\ref{coa1}) and  (\ref{ss1}), we can estimate the values
of the coefficients and observables for gravitational lensing in the
strong field limit. The numerical values of $\theta_{\infty}$, $s$
and $r_{m}$ (which is related to $\mathcal{R}$ by
$r_m=2.5\log{\mathcal{R}}$) are listed in Table I for the different
values of $j$ and $\rho_0/\rho_M$. The dependence of these
observables on the parameters $j$ and $\rho_0/\rho_M$ are also shown
in Figs.(5)-(7).
\begin{table}[h]
\begin{center}
\begin{tabular}{c|cccc|cccc|cccc}
\hline\hline  &&&&&&&&&&&&\\
&\multicolumn{4}{c|}{$\theta_{\infty}$($\mu$
arcsec)}& \multicolumn{4}{c|}{$s$($\mu$ arcsec)}&\multicolumn{4}{c}{$r_m$(magnitudes)} \\
\cline{2-13}&&&&&&&&&&&&\\$\rho_0/\rho_M$&$j=0$&$j=0.03$&$j=0.06$
&$j=0.09$&$j=0$&$j=0.03$&$j=0.06$ &$j=0.09$&$j=0$&$j=0.03$&$j=0.06$&$j=0.09$\\
\hline
&&&&&&&&&&&&\\
0& 26.510&25.955&24.477&22.482&0.0338& 0.0346 &0.0395&0.0493 &6.8219& 6.8219 &6.8219&6.8219\\
0.1&27.374&26.738&25.045&22.765&0.0369& 0.0387 &0.0450&0.0582 &6.7497& 6.7485 &6.7451&6.7398\\
0.2&28.205&27.483&25.565&22.990&0.0407& 0.0430 &0.0509&0.0685 &6.6838& 6.6812 &6.6738&6.6620\\
0.3& 29.005&28.194&26.042&23.162&0.0445&0.0474 &0.0574&0.0803 &6.6234& 6.6193 &6.6070&6.5873\\
0.4&29.779&28.875&26.481&23.285&0.0485& 0.0520 &0.0644&0.0941 &6.5678& 6.5617 &6.5440&6.5149\\
\hline\hline
\end{tabular}
\end{center}
\label{tab1} \caption{Numerical estimation for main observables and
the strong field limit coefficients for the black hole at the center
of our Galaxy, which is hypothesized to be described by the squashed
Kaluza-Klein G\"{o}del black hole spacetime.
$r_m=2.5\log{\mathcal{R}}$.}
\end{table}
From Table I and Fig. (5)-(7), we find that for fixed $j$ with the
increase of $\rho_0$, the angular position of the relativistic
images $\theta_{\infty}$ and the angular separation $s$ increase,
while the relative magnitudes $r_m$ decrease, which is similar to
those in usual squashed Kaluza-Klein black hole spacetime
\cite{schen}. For fixed $\rho_0/\rho_M$, one can obtain that with
the increase of $j$, both $\theta_{\infty}$ and $r_m$ decrease, but
the quantity $s$ increases. These information could help us to
detect the rotation of the cosmological background in the future.

We now make a comparison between the strong gravitational lensing in
the squashed Kaluza-Klein G\"{o}del and four-dimensional Kerr black
hole spacetimes. It is well known that for the photon moving along
the equatorial plane in the four-dimensional Kerr black hole
spacetime \cite{Bozza3,Bozza4} the photon sphere radius, the minimum
impact parameter, the angular position of the relativistic images
$\theta_{\infty}$ and the relative magnitudes $r_m$ in strong
gravitational lensing decreases with the rotation parameter $a$ of
the black hole, while the coefficient $\bar{a}$, the deflection
angle $\alpha(\theta)$ and the observable $s$ in strong
gravitational lensing decrease  with $a$.  These effects of $a$ on
the strong gravitational lensing in the Kerr black hole are similar
to those of the rotation parameter $j$ of cosmological background in
the squashed Kaluza-Klein G\"{o}del case, which can be
understandable since both of  $j$ and $a$ are the rotation
parameters. However, the strong gravitational lensing in the
squashed Kaluza-Klein G\"{o}del black hole spacetime has some
distinct behaviors from that in the Kerr case. In the squashed
Kaluza-Klein G\"{o}del black hole spacetime, the photon sphere
radius, the minimum impact parameter, the coefficient $\bar{a}$,
$\bar{b}$ and the deflection angle $\alpha(\theta)$ in the $\phi$
direction are independent of whether the photon goes with or against
the global rotation of the G\"{o}del Universe. While in the Kerr
black hole, the values of these quantities for the prograde photons
are different from those for the retrograde photons. Moreover, the
coefficient of $\bar{b}$ increases with $j$ in the squashed
Kaluza-Klein G\"{o}del black hole, but decreases with $a$ in the
Kerr case. In the extremely squashed case $\rho_0=0$, we also find
that the coefficient $\bar{a}$ is a constant $1$ and is independent
of the global rotation of the G\"{o}del Universe. These information
could help us to understand further the difference between the
global rotation of the cosmological background and the rotation of
the black hole itself.

\section{summary}

We have investigated the strong gravitational lensing in the neutral
squashed Kaluza-Klein black holes immersed in a rotating
cosmological background. Besides the influence due to the
compactness of the extra dimension, we have disclosed the
cosmological rotational effects in the radius of the photon sphere
and the deflection angle in the $\phi$ direction. For fixed
$\rho_0$, the radius of the photon sphere $\rho_{ps}$ decreases
monotonically with the increase of the G\"{o}del parameter $j$. As
the extra dimension scale $\rho_0$ increases,  $\rho_{ps}$ increases
for the smaller $j$ and decreases for the larger $j$. Moreover, we
also find that the larger values of the parameters $j$ and $\rho_0$
lead to the bigger deflection angle in the $\phi$ direction in the
strong gravitational lensing for the light ray propagating in the
squashed Kaluza-Klein G\"{o}del black hole spacetime. Our result
also show that in the squashed Kaluza-Klein G\"{o}del black hole,
the photon sphere radius, the minimum impact parameter, the
coefficients $\bar{a}$, $\bar{b}$ and the deflection angle
$\alpha(\theta)$ in the $\phi$ direction in the strong gravitational
lensing and the corresponding observables are independent of whether
the photon goes with or against the global rotation of the G\"{o}del
Universe, which is different from those in the Kerr black hole
spacetime.

Assuming that the gravitational field of the supermassive black hole
in the Galactic center can be described by this metric, we estimated
the numerical values of the coefficients and observables in the
strong gravitational lensing. Our results show that the angular
position of the relativistic images $\theta_{\infty}$ and the
relative magnitudes $r_m$ decrease with the increase of the
parameter $j$. The change of the angular separation $s$ with $j$ is
converse to those of $\theta_{\infty}$ and $r_m$. Comparing those
with the data from the astronomical observations in the future, we
could detect whether our universe is rotating or not.

\begin{acknowledgments}
We thank the referee for his/her quite useful and helpful comments
and suggestions, which help deepen our understanding of the strong
gravitational lensing in a squashed Kaluza-Klein G\"{o}del black
hole. This work was  partially supported by the National Natural
Science Foundation of China under Grant No.10875041,  the Program
for Changjiang Scholars and Innovative Research Team in University
(PCSIRT, No. IRT0964) and the construct program of key disciplines
in Hunan Province. J. Jing's work was partially supported by the
National Natural Science Foundation of China under Grant No.10875040
and No.10935013; 973 Program Grant No. 2010CB833004 and the Hunan
Provincial Natural Science Foundation of China under Grant
No.08JJ3010.
\end{acknowledgments}

\end{document}